\documentclass{aa}
\usepackage{graphicx,natbib,hyperref}
\usepackage[varg]{txfonts}
\usepackage{color}

\newcommand{\Porb}{\ensuremath{P_{\mathrm{orb}}}}
\newcommand{\vsi}{\ensuremath{v\,\sin i}}
\newcommand{\kms}{km\,s$^{-1}$}
\newcommand{\Bm}{\ensuremath{\langle B\rangle}}
\newcommand{\Bz}{\ensuremath{\langle B_z\rangle}}
\newcommand{\Feline}{Fe\,{\sc ii}~$\lambda\,6149.2\,\AA$}
\newcommand{\Zeeman}{\Delta\lambda_{\rm Z}}

\bibpunct[ ]{(}{)}{;}{a}{}{,}

\begin{document}

\title{HD~213258: a new rapidly oscillating, super-slowly rotating, strongly magnetic Ap star in a spectroscopic binary}

\author{G.~Mathys\inst{1}
    \and V.~Khalack\inst{2}
	\and O.~Kobzar\inst{2}
	\and F.~LeBlanc\inst{2}
        \and P.~L.~North\inst{3}
      }

\institute{European Southern Observatory,
  Alonso de Cordova 3107, Vitacura, Santiago, Chile\\\email{gmathys@eso.org}
  \and
  D\'epartement de Physique et d'Astronomie, Universit\'e de Moncton,
  Moncton, NB, Canada E1A 3E9
  \and
  Institut de Physique, Laboratoire d'astrophysique, Ecole
  Polytechnique F\'ed\'erale de Lausanne (EPFL), Observatoire de
  Sauverny, CH-1290 Versoix, Switzerland}

\date{Received $\ldots$ / Accepted $\ldots$}

\titlerunning{HD~213258}

\abstract{We report about HD~213258, an Ap star that we recently
  identified as presenting a unique combination of rare, remarkable
  properties. Our study of this star is based on ESPaDOnS Stokes $I$
  and $V$ data obtained at 7 epochs spanning a time interval slightly
  shorter than 2 years, on TESS data, and on radial velocity
  measurements from the CORAVEL data base. We confirm that HD~213258,
  which was originally
  suspected to be a F\,str\,$\lambda4077$ star, is definitely an Ap
  star. We found that, in its spectrum,
  the \Feline\ line is resolved
  into its two magnetically split components. The mean magnetic field
  modulus of HD~213258, $\Bm\sim3.8$\,kG, which we determined from
  this splitting, does not show significant variations over
  $\sim$2~years. Comparing our mean longitudinal field determinations
  with a couple of measurements from the literature, we show that the
  stellar rotation period must likely be of the order of 50 years,
  with a reversal of the field polarity. Moreover, HD~213258 is a
  rapidly oscillating Ap (roAp) star, in which high overtone
  pulsations with  
  a  period of 7.58\,min are detected.
  Finally, we confirm that HD~213258 has a mean radial
  velocity exceeding (in absolute value) that of at least 99\% of the
  Ap stars. The radial velocity shows low amplitude variations, which
  suggests that the star is a single-line spectroscopic binary. It is
  also a known astrometric binary. While its orbital
  elements remain to be determined, its orbital period likely is one
  of the shortest known for a binary roAp star.
Its secondary is close to the borderline between stellar and
  substellar objects. There is a significant probability that it may be
  a brown dwarf. The pair represents a
  combination that has never been observed before. While most
of the  
  above-mentioned properties, taken in isolation, are observed in a
  small fraction of the whole population of Ap stars, the probability
  that a single star possesses all of them is extremely low. This
  makes HD~213258 an exceptionally interesting object that deserves to
  be studied in detail in the future.}


\maketitle

\section{Introduction}
\label{sec:intro}
The peculiarity of HD~213258 (= BD+35~4815) was first reported by
\citet{1985AJ.....90..341B}. He assigned it to a new group of upper
main-sequence chemically peculiar (CP) stars that he had recently
identified \citep{1981AJ.....86..553B}: the F\,str\,$\lambda4077$
stars. Originally, this classification referred to stars whose spectra
resemble those of the Am stars, but show an abnormally strong
Sr\,{\sc ii}~$\lambda\,4077$\,\AA\ line. However, as noted by
\citet{1987A&A...186..191N}, a specific definition of the criteria
that allow one to distinguish the F\,str\,$\lambda4077$ stars from the
Am, Ap Sr or $\lambda$~Boo stars was missing at the
time.\footnote{Later, \citet{1991A&A...244..335N} found that at least
  half of the F\,str\,$\lambda4077$ stars are main sequence Ba stars,
  owing their chemical peculiarity to binary evolution.} This left
some ambiguity in the classification. As a matter of fact, in their
Table~3, \citet{1991A&A...244..335N} flagged HD~213258 as a possible
Ap star. Nevertheless, it was not included in the {\em Catalogue of
  Ap, HgMn and Am stars\/} \citep{2009A&A...498..961R}.

In this short note, we report that HD~213258 has a quasi-unique
combination of rare, remarkable properties, with respect to its
magnetic field (Sect.~\ref{sec:bfield}), its rotation
(Sect.~\ref{sec:rot}), its space velocity and its binarity
(Sect.~\ref{sec:rv}), and its pulsation (Sect.~\ref{sec:puls}). As a
conclusion, in Sect.~\ref{sec:conc}, we show why this combination of
properties makes HD~213258 an object of exceptional interest that
deserves to be studied in detail in the future.

\begin{figure*}
  \resizebox{\hsize}{!}{\includegraphics{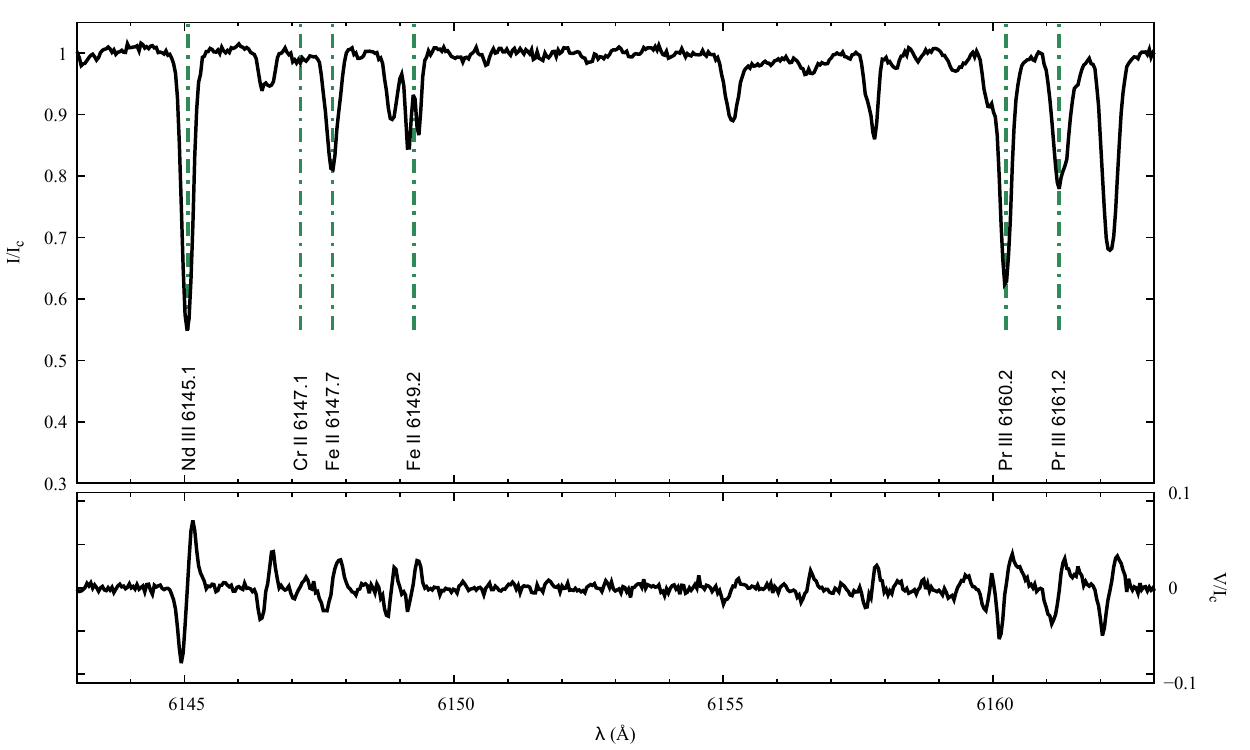}}
  \caption{Portion of the spectrum of HD~213258 recorded on
      HJD\,2,459,180.728 in Stokes $I$
    (\textit{top\/}) and $V$ (\textit{bottom\/}), showing the resolved
  magnetically split line \Feline. A few other lines that are
  typical of Ap stars are identified. The spectrum has been normalised
  to the continuum ($I_{\rm c}$) and the wavelength scale has been
  converted to the laboratory reference frame.}
\label{fig:spectrum}
\end{figure*}

\section{Magnetic field}
\label{sec:bfield}
The ESPaDOnS spectrograph at the Canada-France-Hawaii Telescope (CFHT)
was used to record Stokes $I$ and $V$ spectra of HD~213258 at seven
epochs between November 2020 and October 2022. They cover the spectral
range 3700--10000\,\AA, at a resolving power $R\sim65,000$. They were
reduced by the CFHT team using the dedicated software package
Libre-ESPrIT \citep{1997MNRAS.291..658D}. The resulting S/N ratio in
Stokes $I$  reaches its maximum of about
400 or more in echelle order \#32 (7080\,\AA). A portion
of one of these spectra is shown for illustration in
Fig~\ref{fig:spectrum}. The sharpness of the spectral lines is
striking. One can see that the \Feline\ line is resolved into its magnetically
split components. This is indicative of a very low projected equatorial
velocity $\vsi$ and of the presence of a strong magnetic field. The
quantitative determination of the latter is discussed below.

One can also note in the spectrum the presence of strong lines that
are characteristic of typical Ap stars, such as Nd\,{\sc
  iii}~$\lambda\,6145.1$\,\AA, Cr\,{\sc ii}~$\lambda\,6147.1$\,\AA,
Pr\,{\sc iii}~$\lambda\,6160.2$\,\AA\ and Pr\,{\sc
  iii}~$\lambda\,6161.2$\,\AA. The presence of these lines
resolves the ambiguity that affected the original
classification of HD~213258 and indicates that it is definitely an Ap
star.

On the other hand, all the lines present in the observed spectrum show
clear Stokes $V$ signatures. They reveal the presence of a sizeable
component of the magnetic field along the line of sight, which does not
average out over the stellar hemisphere that was visible at the epoch
of the observation.

The mean magnetic field modulus \Bm, defined
as the line-intensity
weighted average over the visible stellar hemisphere of the modulus of
the magnetic vector, was determined from the wavelength separation of
the resolved components of the \Feline\
line. The magnetic splitting pattern of this line is a doublet. 
The value of $\Bm$ is derived by application of the following formula:
\begin{equation}
\lambda_{\rm r}-\lambda_{\rm b}=g\,\Zeeman\,\Bm\,.
\label{eq:Bm}
\end{equation}
In this equation, $\lambda_{\rm r}$ and $\lambda_{\rm b}$ are,
respectively, the wavelengths of the red and blue split line
components; $g$ is the Land\'e factor of the split level of the
transition ($g=2.70$; \citealt{1985aeli.book.....S});
$\Zeeman=k\,\lambda_0^2$, with
$k=4.67\,10^{-13}$\,\AA$^{-1}$\,G$^{-1}$; $\lambda_0=6149.258$\,\AA\
is the nominal wavelength of the considered transition.

The procedure used to measure the wavelengths $\lambda_{\rm b}$ and
$\lambda_{\rm r}$ of the \Feline\ split line
components has been described i n detail by \citet{1992A&A...256..169M}
and by \citet{1997A&AS..123..353M}. As in many other Ap stars, the
\Feline\ in HD~213258 is blended on the blue side with an unidentified
rare-earth line. We disentangled the contribution of the latter from
that of the two \Feline\ line components by fitting three
Gaussians to them. As shown by \citet{1997A&AS..123..353M}, this
represents a very effective way to achieve consistent determinations
of the wavelengths $\lambda_{\rm b}$ and $\lambda_{\rm r}$, hence of
the value of $\Bm$.

The difficulty in estimating the uncertainty affecting the derived
values of the mean magnetic
field modulus was discussed in detail by
\cite{1997A&AS..123..353M}. In the present case, since the
measurements obtained until now sample only a fraction of the stellar
rotation cycle (see 
Sect.~\ref{sec:rot}), we follow the prescription of these authors and
estimate the 
uncertainty of the \Bm\ determinations in HD~213258 to be of the order
of 40\,G, from comparison of the
profile of the \Feline\ line in HD~213258 with its profile in other
stars for which this uncertainty is better constrained.

The mean longitudinal magnetic field \Bz\ is the line-intensity
weighted average over the visible stellar hemisphere of the component
of the magnetic vector along the line of sight. It is determined from
the wavelength shift of the spectral lines between the two circular
polarisations by application of the following formula:
\begin{equation}
\lambda_{\rm R}-\lambda_{\rm L}=2\,\bar g\,\Zeeman\,\Bz\,,
\label{eq:Bz}
\end{equation}
where $\lambda_{\rm R}$ (respectively $\lambda_{\rm L}$) is the wavelength of
the centre of gravity of the line in right (respectively left) circular
polarisation and $\bar g$ is the effective Land\'e factor of the
transition. $\Bz$ is determined through a
least-squares fit of the measured values of $\lambda_{\rm
  R}-\lambda_{\rm L}$ by a function of the form given above. The standard error
$\sigma_z$ that is derived from that
least-squares analysis is used as an estimate of the uncertainty
of the obtained value of $\Bz$.

The results of the magnetic measurements of HD~213258 are presented in
Table~\ref{tab:Bmeas}. Column~1 gives the Heliocentric Julian Date of
mid-exposure for each of the analysed spectra.
The mean magnetic field
modulus values are listed in Col.~2. Columns~3 to 5 present the
results of the determination of the mean longitudinal magnetic field:
its value $\Bz$, its uncertainty $\sigma_z$ and the number $N$ of
diagnostic lines that were measured. These are lines of Fe\,{\sc i},
which span the wavelength range 4175--6180\,\AA. The same lines were
also used to derive the radial velocity of HD~213358, from a
least-squares fit of the wavelength shifts of the centres of gravity of the
Stokes~$I$ line profiles with respect to their laboratory values,
against these laboratory values,
using the same diagnostic lines as
for the $\Bz$ determinations. The derived values of the heliocentric radial
velocity $v_{\rm r}$ and their uncertainties $\sigma_v$ (that is, the
standard error of the least-squares analysis) are given in Cols.~6 and
7 of Table ~\ref{tab:Bmeas}.

\begin{figure}
  \resizebox{\hsize}{!}{\includegraphics{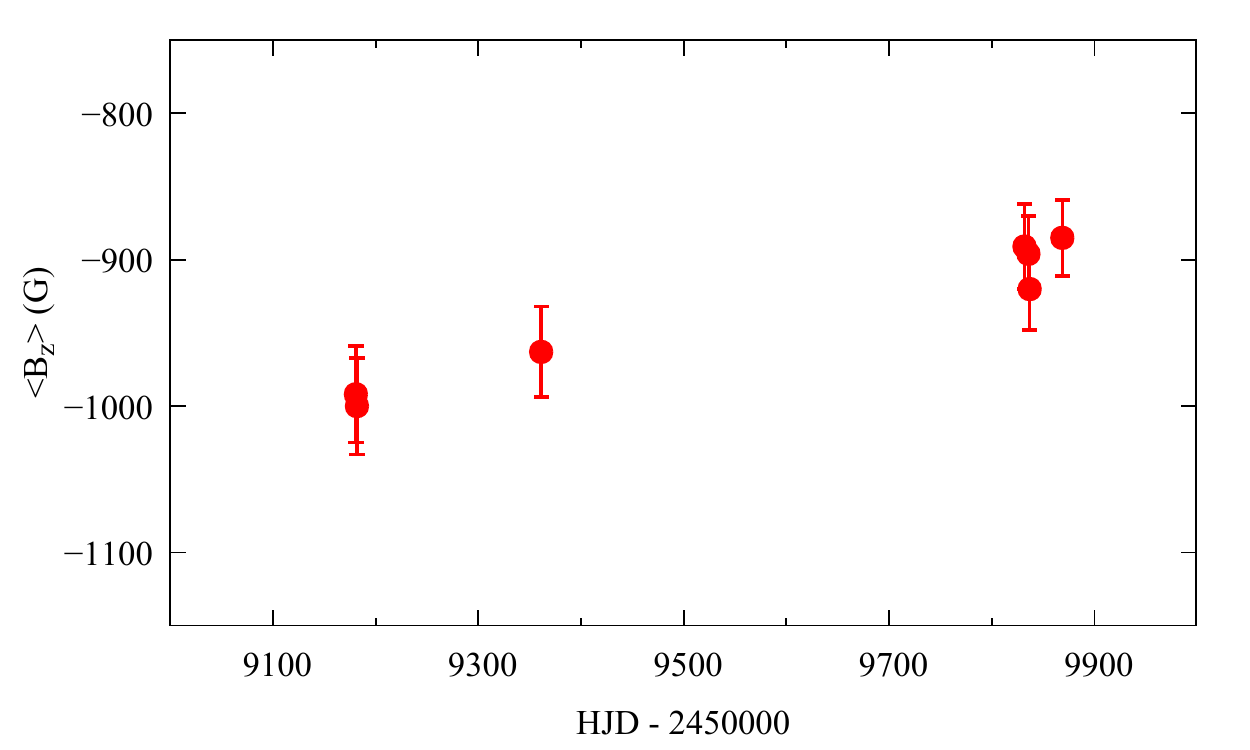}}
  \resizebox{\hsize}{!}{\includegraphics{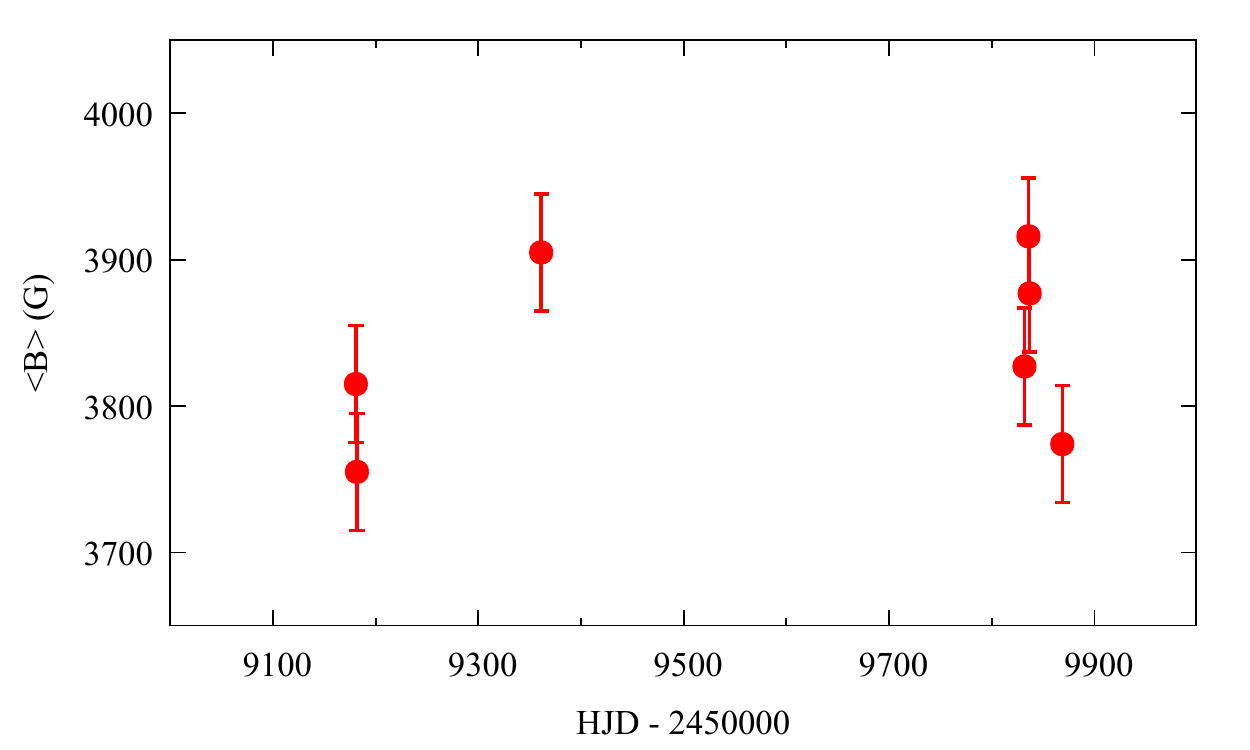}}
  \caption{Mean longitudinal magnetic field (\textit{top\/}) and mean
    magnetic field modulus (\textit{bottom\/}) of HD~213258 against
    time.}
  \label{fig:bfield}
\end{figure}

\section{Rotation}
\label{sec:rot}
The magnetic measurements of Sect.~\ref{sec:bfield} are plotted
against time in Fig.~\ref{fig:bfield}. In the upper panel, the mean
longitudinal magnetic field shows an apparently linear trend from the
first epoch of observation to the most recent one, from more negative
to less negative values. All \Bz\ values are comprised in a narrow
range, between $-1000$\,G and $-885$\,G. However, two earlier mean
longitudinal magnetic field determinations from the literature,
performed by \citet{1992pess.conf...51N} using spectra recorded with
the Zeeman analyser of the Main Stellar Spectrograph of the 6\,m BTA
telescope of the Special Astrophysical Observatory
\citep{2014AstBu..69..339P} on 
August 12 and 13, 1989, yielded respectively $\Bz=-198\pm133$\,G and 
$\Bz=210\pm108$\,G. Given the uncertainties affecting them, these
measurements do not significantly differ from each other, nor from
zero. They represent a strong indication that the overall amplitude of
variation of $\Bz$ in HD~213258 is considerably greater than observed
over $\sim$2~years with ESPaDOnS, 
hence that the stellar rotation
period must vastly exceed two years. Thus HD~213258 is a newly
identified member of the group of the super-slowly rotating Ap (ssrAp)
stars \citep{2020pase.conf...35M}.

The magnetic measurements obtained until now sample the rotation
  cycle too sparsely to constrain the shapes of the variation curves of
  either \Bm\ or \Bz. Even trying to characterise these shapes under
  the frequently made assumption of a magnetic field structure that is
  to first order dipolar would represent an overinterpretation of the
  available data. Therefore, the discussion below is based on a linear
approximation, which is sufficient to set meaningful constraints on
the lower limit of the rotation period. Indeed, from one extremum to
the next, and away from 
both extrema, a straight line does not drastically depart from the
actual sinusoidal shape of the $\Bz$ variation curve corresponding to a
dipole.

Extrapolating the linear trend of variation of the mean longitudinal
magnetic field observed over the past two years (690 days), it would
take HD~213258 of the order of 5060~d from the most recent ESPaDOnS
observation to get back to a value of $\Bz\sim0$\,G similar to that
derived in 1989. This suggests a minimum value of $\sim$5750\,d for
half the rotation period. The full rotation period should be at least
twice as long, that is, have a minimum value of about 11,500\,d, or
31.5\,yr. However, the time elapsed between the observations of
\citet{1992pess.conf...51N} and the first of the present ESPaDOnS
observation is 11,432\,d. These observations were definitely obtained
at very different rotation phases, so that they are inconsistent with
values of the rotation period close to 11,500\,d. The actual period
must be much longer. Accordingly, the $\Bz$ values from 1989 and 2020
cannot both be close to the extrema of the mean longitudinal magnetic
field. Either the negative $\Bz$ extremum must have an absolute value
greater than 1\,kG, or $\Bz$ must become positive over part of the
rotation cycle, and reach a positive extremum, or both extrema must be
outside the range of $\Bz$ values that have been observed until
now. For instance, if $\Bz$ was close to its negative extremum in
2020, and if its variation curve is approximately symmetric about
this extremum, the mean longitudinal magnetic field could plausibly
have been positive between $\sim$1989 and $\sim$2002, and it could
have gone through its positive extremum around 1995 or 1996. This
suggests that the minimum value of the rotation period should be at
least of the order of 50\,yr. The period might even be a few years
longer if over part of it $\Bz$ reached values more negative than
$-1.0$\,kG. While consideration of the mean magnetic field modulus
sets an upper limit to the maximum absolute value of the mean
longitudinal field, comparison with other stars, such as HD~93507
\citep{2017A&A...601A..14M}, shows that values as negative as
$\Bz=-1.25$\,kG could be reached by HD~213258.

\begin{table}
  \caption{Mean magnetic field modulus, mean longitudinal magnetic
    field and heliocentric radial velocity measurements.}
  \label{tab:Bmeas}
  \centering
\small{\begin{tabular}{ccccccc}
    \hline\hline\\[-4pt]
    HJD&$\Bm$&$\Bz$&$\sigma_z$&$N$&$v_{\rm r}$&$\sigma_{\rm r}$\\
       &(G)&(G)&(G)&&(\kms)&(\kms)\\[4pt]
    \hline\\[-4pt]
    2459180.728&3815&$-$992&33&50&$-$87.64&0.15\\
    2459181.714&3755&\llap{$-$}1000&33&50&$-$87.76&0.15\\
    2459361.124&3905&$-$963&31&50&$-$88.97&0.15\\
    2459832.014&3827&$-$891&29&50&$-$87.91&0.13\\
    2459835.847&3916&$-$896&26&50&$-$87.95&0.14\\
    2459836.993&3877&$-$920&28&49&$-$87.55&0.11\\
    2459868.858&3774&$-$885&26&50&$-$87.85&0.13\\[4pt]
    \hline
  \end{tabular}}
\end{table}

\begin{figure}
  \resizebox{\hsize}{!}{\includegraphics{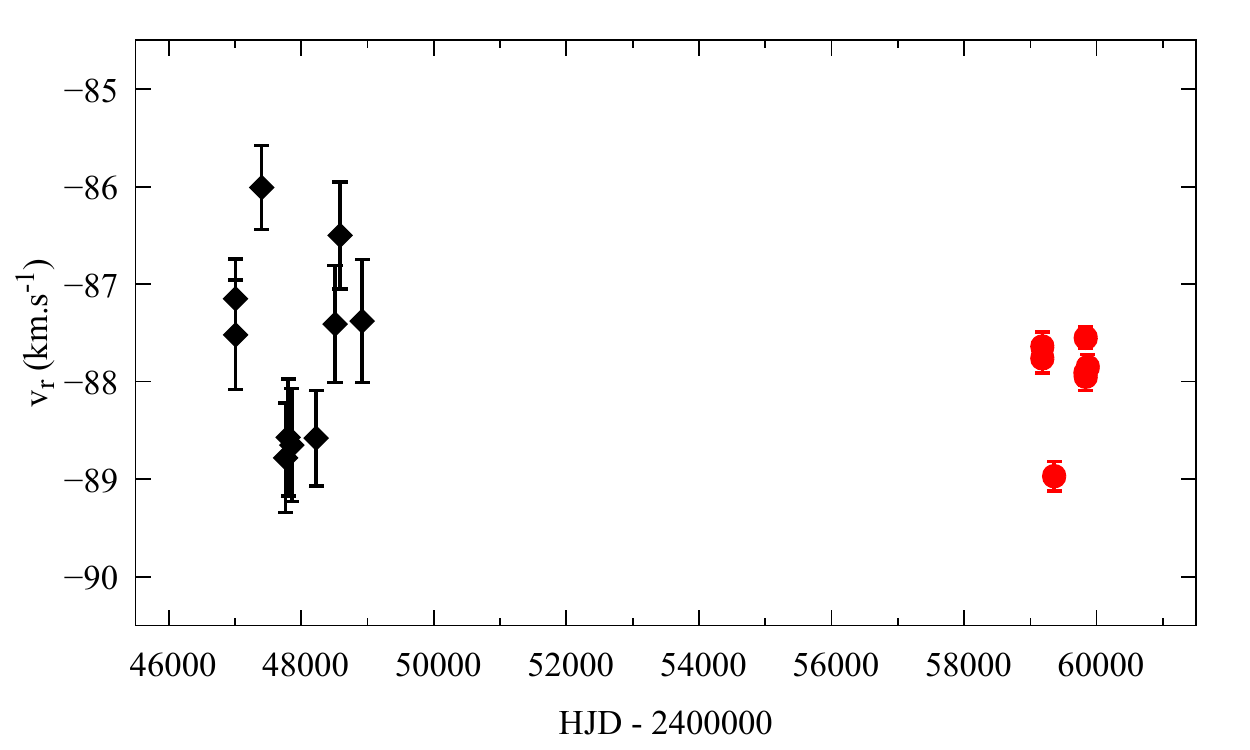}}
  \caption{Heliocentric radial velocity of HD~213258 against
    time. \textit{Black diamonds\/} identify the measurements based on
    CORAVEL observations; \textit{red dots\/} those based on
    ESPaDOnS spectra.}
  \label{fig:rv}
\end{figure}

The lower panel of Fig.~\ref{fig:bfield} show the measurements of the
mean magnetic field modulus of HD~213258. This field moment does not
show any significant variation over the ~2-yr time interval spanned by
the ESPaDonS spectra. This is not unexpected, as 2 years probably
represent less than 0.04 rotation period. The relative amplitude
of the variations of \Bm\ does not exceed 30\% in most Ap stars
\citep{2017A&A...601A..14M}, so
that such variations may not be detectable over too small a fraction
of a rotation cycle.

\section{Radial velocity and binarity}
\label{sec:rv}
The spectral line measurements carried out to determine the mean
longitudinal magnetic field were used to derive the heliocentric
radial velocity of
HD~213258 at the seven epochs of observation with ESPaDOnS. The
average of the seven individual values, $-87.95$\,\kms, is consistent
with the average radial velocity computed by
\citet{1991A&A...244..335N} from six CORAVEL measurements,
$-86.8$\,\kms. These six
CORAVEL values, together with four later ones, are listed in
Table~\ref{tab:rvmeas}. All the radial velocity values of
Tables~\ref{tab:Bmeas} and \ref{tab:rvmeas} are also plotted
in Fig.~\ref{fig:rv}. The CORAVEL and ESPaDOnS data sets appear
mutually consistent, in line with the conclusion reached by
\citet{2017A&A...601A..14M} for other stars observed with these two
instruments. We also confirm the claim by \citet{1991A&A...244..335N}
that the radial velocity of HD~213258 is definitely variable. In
particular, the difference between the measurement obtained on
JD~2459361 and the six other contemporaneous radial velocity
determinations, although small (of the order of 1\,\kms), is highly
significant. \citet{1991A&A...244..335N} contemplated the possibility
that the radial velocity variations that they detected could reflect
the changing aspect of the visible hemisphere of a spotted star over
its rotation period. This interpretation can be ruled out given the
above-mentioned strong evidence that the rotation period of the star
is of the order of decades. Thus, HD~213258 must almost certainly be
part of a single-lined spectroscopic binary system. Indeed, no obvious
evidence of the secondary was found in the spectra.

\begin{table}
  \caption{Heliocentric radial velocity measurements obtained with CORAVEL.}
  \label{tab:rvmeas}
  \centering
  \begin{tabular}{ccc}
    \hline\hline\\[-4pt]
    HJD&$v_{\rm r}$&$\sigma_r$\\
       &(\kms)&(\kms)\\[4pt]
    \hline\\[-4pt]
     2447006.619&$-$87.15&0.41\\
     2447007.598&$-$87.52&0.56\\
     2447401.503&$-$86.01&0.43\\
     2447762.523&$-$88.78&0.56\\
     2447801.446&$-$88.57&0.60\\
     2447856.353&$-$88.65&0.58\\
     2448223.251&$-$88.58&0.49\\
     2448510.469&$-$87.41&0.60\\
     2448585.308&$-$86.50&0.55\\
     2448917.417&$-$87.38&0.63\\[4pt]
    \hline
  \end{tabular}
\end{table}

There are too few radial
velocity measurements and the epochs of CORAVEL and ESPaDOnS
observations are too far apart from each other to allow the orbital
elements to be 
determined. Consideration of Fig.~\ref{fig:rv} suggests that the
orbital period may be of the order of weeks, or possibly
longer. Indeed, radial velocity values determined from
  observations obtained on 2--3 consecutive or nearly consecutive
  nights do 
  not show any significant variations, but variations occur between
  such groups of observations spaced from each other by a few
  weeks. In any case, the 
  orbital period of HD~213258 must be much shorter than its rotational
  period. This occurs frequently for ssrAp stars in binaries
  \citep{2017A&A...601A..14M}. More observations are needed to
characterise the system better.

Furthermore, from analysis of the proper motion anomaly between
  the \textsc{Hipparcos} catalogue and  the
  Early Third Data Release (EDR3) of the Gaia
  catalogue, \citet{2021ApJS..254...42B} and
  \citet{2022A&A...657A...7K} also showed that 
  HD~213258 is an astrometic binary.
  Assuming a circular orbit observed face-on and using a mass estimate
  $m_1=1.70\,M_{\sun}$ for the Ap primary, \citet{2022A&A...657A...7K}
  derived estimates of the mass $m_2$ of the secondary:
  $m_2=129.33\,M_J$ for an orbital radius $r=3$\,au (corresponding to
  an orbital period $\Porb\sim4.0$\,yr), $m_2=69.56\,M_J$ for
  $r=5$\,au ($\Porb\sim11.2$\,yr), and $m_2=87.97\,M_J$ for $r=10$\,au
  ($\Porb\sim31.6$\,yr). This puts the secondary of HD~213258 close to
  the borderline between stellar and substellar objects. There is a
  significant probability that it may be a brown dwarf.
    
The average value of the heliocentric radial velocity of HD~213258, of
the order of $-88$\,\kms, is exceptionally large (in absolute value)
for an Ap star. For instance, in the systematic study of a sample of 186
Ap stars by \citet{1996A&AS..118..231L}, the values of the
heliocentric radial velocity range from $-40$ to $42$\,\kms. However,
as noted by \citet{1991A&A...244..335N}, the space velocity of
HD~213258 is essentially radial, so that while large, it is not
extreme. We are not aware of any modern study of the space
  velocity distribution of Ap stars, so that we cannot at present put
  the high radial velocity of HD~213258 in perspective, nor understand
  its implications (if any) on the other stellar properties.

\section{Pulsation}
\label{sec:puls}

\begin{figure}
  \resizebox{\hsize}{!}{\includegraphics{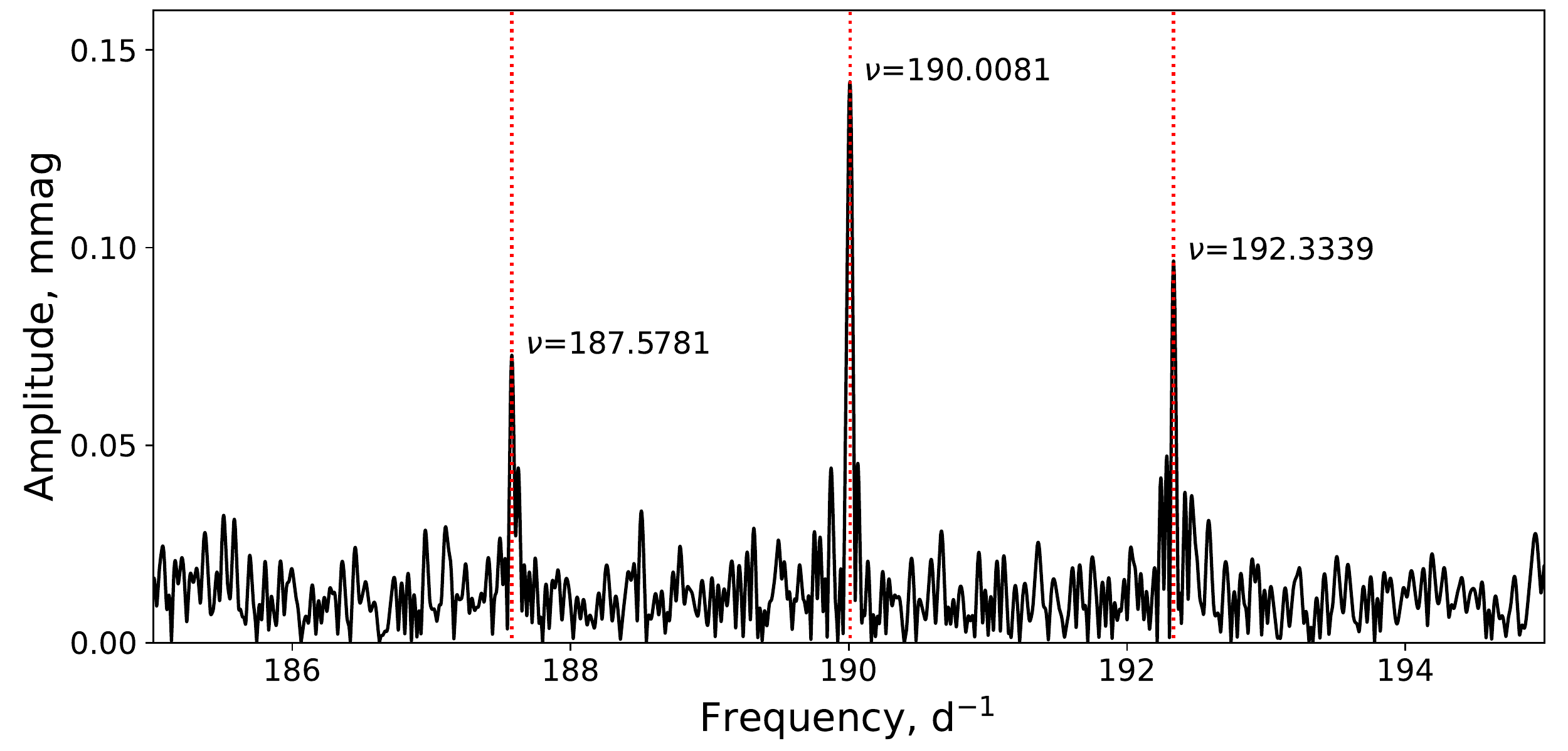}}
  \caption{High-overtone pulsations of HD~213258 detected during the analysis of photometric data in sector~56 provided by \textit{TESS}. \textit{Vertical red dotted lines\/} specify the position of three significant signals in the Lomb-Scargle periodogram.}
  \label{fig:puls}
\end{figure}

The star HD~213258 has been observed twice, in Sectors~16 and 56, by the
space telescope TESS during the first 4 years of its
operation. Employing the utilities of the Lightkurve Python package
designed for analysis of Kepler and TESS data
\citep{2018ascl.soft12013L} and a Python code developed by Jonathan
Labadie-Bartz (2022, private communication),  the TESS images of this
target observed in Sector~56 with a cadence of 158\,s were
downloaded from the Mikulski Archive for Space Telescopes
(MAST)\footnote{https://archive.stsci.edu/}. The light curve was
extracted from the TESS Full Frame Images in a manner similar to 
\citet{2022AJ....163..226L}. The cut-out images with a size of
24$\times$24 pixels were used to infer the raw light curve of
HD~213258 and to remove the sky background 
using the Principal Component Analysis (PCA) detrending method. 
The derived flux was transformed to stellar magnitudes and analysed
for periodic variability using a Lomb-Scargle periodogram
\citep{2018ApJS..236...16V}. 

The Lomb-Scargle periodogram clearly shows a distinguishable triplet
at frequencies around 190\,d$^{-1}$ (see Fig.~\ref{fig:puls}). The
central mode has the highest amplitude  
and corresponds to the period $P=7.579(2)$\,min. There are three
frequencies with significant amplitudes that are split with an average
frequency separation of $\delta\nu=2.38(5)$\,d$^{-1}$. This kind of
high-overtone pulsations is typically observed in roAp stars
\citep{1982MNRAS.200..807K}. The detection of roAp type pulsations and
a significant magnetic field in HD~213258 (see Fig.~\ref{fig:bfield})
are strong proofs that it is a chemically peculiar roAp star with 
extremely slow rotation (see Sect.~\ref{sec:rot}). 

\section{Conclusion}
\label{sec:conc}
The Ap star HD~213258 presents a quasi-unique combination of
remarkable properties, each of which is observed in isolation in only
a small fraction of the entire population of the Ap stars.  A few
percent of these stars have rotation periods longer than 1\,yr
\citep{2017A&A...601A..14M,2018MNRAS.475.5144S}. The longest period
that has been accurately determined for an Ap star is that of
HD~50169, which is 29 years long \citep{2019A&A...624A..32M}. The only
Ap star for which a lower limit of the period value greater than
50\,yr\ has been set until now is $\gamma$~Equ
\citep{2016MNRAS.455.2567B}. If HD~213258 has a rotation period of the
order of 50\,yr, as we advocate, it is one of the most promising
candidates for detailed study of extreme slow rotation in Ap stars.

The lack of rotational broadening of the spectral lines of the ssrAp
stars lends itself well to the resolution of 
magnetically split lines. But the number of stars showing such
resolution remains small, of the order of a few percent of all Ap
stars \citep{2017A&A...601A..14M}, as the threshold of detection of
magnetically resolved lines in the visible is of the order of 2\,kG
and magnetic fields of several kG are rare. Thus, the fact that the
\Feline\ line is resolved into its magnetic components in HD~213258 is
a distinctive trait.

While the rate of occurrence of roAp stars among ssrAp stars, which
may reach $\sim$20\%, is considerably higher than the fraction of roAp
stars among all Ap stars \citep{2022A&A...660A..70M}, roAp stars
seldom belong to binary systems
\citep{2000A&A...355.1031H,2012A&A...545A..38S,2019MNRAS.488...18H}.
Such binary systems are wide: the shortest orbital period that has
been accurately determined for a pair containing an roAp star, HD~42659, is
$93\fd2$ \citep{2015A&A...582A..84H}. The radial velocity measurements
obtained until now for the roAp star HD~213258 (see
Tables~\ref{tab:Bmeas} and 
\ref{tab:rvmeas}) do not rule out the possibility of a shorter
period: it will be very valuable to determine $v_{\rm r}$ at
additional epochs to constrain its orbital elements. Even more
remarkably, there is a significant probability that the secondary of
the pair may be 
substellar. If confirmed, this would make HD~213258 the first roAp 
star known to have a brown dwarf companion. In any event, it is
certainly the roAp star with the least massive companion that is known
to this date.

Furthermore, it is rather infrequent for roAp stars to have very
strong magnetic fields. For instance, of the 44 roAp stars for which
an averaged value of the magnetic field is given in Table~1 of
\citet{2015MNRAS.452.3334S}, only 8 (18\%) have field values greater
than 3.5\,kG. Admittedly, this statistic is very approximative since it is
based on inhomogeneous literature sources and magnetic field
measurements of
differing completeness and quality. But this does not detract from the
fact that, with $\Bm\sim3.8$\,kG, HD~213258 ranks among the most
strongly magnetic roAp stars.

Finally, as illustrated in Sect.~\ref{sec:rv}, the mean radial
velocity of HD~213258 appears greater than that of more than 99\%\ of
the Ap stars.

In summary, each of the above-mentioned properties of HD~213258 taken
in isolation puts it in a minority group of Ap stars that may
represent between $\sim$20\%\ and less than 1\%\ of the whole
population of Ap stars. What makes HD~213258 especially remarkable is
that it presents a combination of all these rare properties. Such a
combination must have a very low probability to occur, and to the best of our
knowledge, HD~213258 is until now the only known Ap star possessing
it. For instance, $\gamma$~Equ is an roAp star that has a
  magnetic field of the same order of magnitude as that of HD~213258
  and a rotation period that is probably longer than that of
  HD~213258. But while the importance of the much higher space
  velocity of the latter is unclear, that its companion is
  much closer than that of $\gamma$~Equ and that
it has a very low mass and may plausibly be
  a brown dwarf mark HD~213258 as really unique. The
purpose of this note is to call the attention of the stellar 
astrophysics community to this star, so that it receives the attention
that it deserves. Indeed, for instance, this star is an excellent
workbench to study the effects of a strong magnetic field and of a
  companion on the atmospheric and pulsation
properties of Ap stars. We are currently working to determine its
fundamental parameters, the chemical abundances in its atmosphere, and
their vertical stratification. The results of this detailed analysis
will be the subject of a future paper.

 \begin{acknowledgements} 
V.K. and F.L. acknowledge support from the Natural Sciences and Engineering Research Council of Canada (NSERC).
O.K. and V.K. are thankful to the Facult\'{e} des \'{E}tudes Sup\'{e}rieures et de la Recherce and to the Facult\'{e} des Sciences de l'Universit\'{e} de Moncton for the financial support of this research.
This paper includes data collected with the TESS mission, obtained from the MAST data archive at the Space Telescope Science Institute (STScI). Funding for the TESS mission is provided by the NASA Explorer Program. STScI is operated by the Association of Universities for Research in Astronomy, Inc., under NASA contract NAS 5–26555.
We thank the TESS and TASC/TASOC teams for their support of the present work. This research has made use of the SIMBAD database, operated at CDS, Strasbourg, France.
Part of the analysed spectra has been obtained with the spectropolaimeter ESPaDOnS at the Canada-France-Hawaii Telescope (CFHT) which is operated by the National Research Council of Canada, the Institut National des Sciences de l'Univers of the Centre National de la Recherche Scientifique of France, and the University of Hawaii.
 \end{acknowledgements} 

\bibliographystyle{aa}

\bibliography{hd213258}

\end{document}